# Orbital torque switching of room temperature two-dimensional van der Waals ferromagnet Fe$_3$GaTe$_2$

Delin Zhang[1,2,3,†,*], Heshuang Wei[1,†], Jinyu Duan[1,†], Jiali Chen[4,†], Dongdong Yue[5], Yuhe Yang[2], Jinlong Gou[1], Junxin Yan[5], Kun Zhai[5], Ping Wang[1], Shuai Hu[1], Zhiyan Jia[2], Wei Jiang[4,*], Wenhong Wang[1,2,3], Yue Li[3], Yong Jiang[1,2,3,*]

1. Institute of Quantum Materials and Devices; School of Electronic and Information Engineering, Tiangong University, Tianjin, 300387, China.
2. School of Material Science and Engineering; State Key Laboratory of Separation Membrane and Membrane Processes, Tiangong University, Tianjin, 300387, China.
3. School of Physical Science & Technology, Tiangong University, Tianjin, 300387, China.
4. Centre for Quantum Physics, Key Laboratory of Advanced Optoelectronic Quantum Architecture and Measurement (MOE); Beijing Key Lab of Nanophotonics & Ultrafine Optoelectronic Systems, School of Physics, Beijing Institute of Technology, Beijing, 100081, China.
5. Center for High Pressure Science, State Key Laboratory of Metastable Materials Science and Technology, Yanshan University, Qinhuangdao 066000, China

†These authors contributed equally: Delin Zhang, Heshuang Wei, Jinyu Duan, Jiali Chen
*Corresponding authors. Email: zhangdelin@tiangong.edu.cn (D.L.Z.), wjiang@bit.edu.cn (W.J.) and yjiang@tiangong.edu.cn (Y.J.)

## Abstract

**Efficiently manipulating the magnetization of van der Waals (vdW) ferromagnets has attracted considerable interest in developing room-temperature two-dimensional (2D) material-based memory and logic devices. Here, taking advantage of the unique properties of the vdW ferromagnet as well as promising characteristics of the orbital Hall effect, we demonstrate the room-temperature magnetization switching of vdW ferromagnet Fe$_3$GaTe$_2$ through the orbital torque generated by the orbital Hall material, Titanium (Ti). The switching current density is estimated to be around 1.6×10$^6$ A/cm$^2$, comparable to that achieved in Fe$_3$GaTe$_2$ using spin-orbit torque from spin Hall materials (e.g., WTe$_2$, and TaIrTe$_4$). The efficient magnetization switching arises from the combined effects of the large orbital Hall conductivity of Ti and the strong spin-orbit correlation of the Fe$_3$GaTe$_2$, as confirmed through theoretical calculations. Our findings advance the understanding of orbital torque switching and pave the way for exploring 2D material-based orbitronic devices.**

Spintronic devices have been extensively studied for their potential in developing energy-efficient memory and computing components, offering ultrahigh storage density, ultrafast switching speed, ultralow energy consumption, and excellent scalability (1-3). These devices are primarily manipulated through spin-transfer torque (STT), where the charge current can be converted into spin current via the ferromagnets with high spin polarization (4), or by spin-orbit torque (SOT), where the conversion happens through spin Hall materials (SHMs) with strong spin-orbit coupling (SOC) (5-7). Despite significant advancements over the past decade, spintronic memory devices still face several challenges. STT-based memory devices struggle with relatively low endurance, high switching current density, and Joule heating caused by current flowing through the tunnel barrier. SOT-based devices require SHMs with strong SOC for efficient spin-torque generation. Recently, attention has increasingly shifted toward orbitronic devices, where device manipulation is driven by orbital torque (OT) generated by the orbital Hall effect (OHE) through an orbital Hall material (OHM) with the weak SOC (8-22). This approach has the potential to overcome the challenges faced by spintronic devices.

The OHE originates from the orbital texture created by orbital hybridization, which generates finite orbital angular momentum along the direction of $\mathbf{E} \times \mathbf{k}$ under an external electric field ($\mathbf{E}$) (8-10). Unlike STT (4) and SOT devices, which rely solely on charge-to-spin conversion (5-7), OT devices involve two stages of current conversion: the charge-to-orbital current conversion in the OHM and the orbital-to-spin current conversion in the adjacent ferromagnetic (FM) material. This unique mechanism allows for tuning the OT efficiency ($\xi_{OT}$) by designing novel heterostructures with optimal selections of OHM and FM material combinations (20). So far, many OHMs have been predicted to efficiently convert charge current to orbital current due to their giant orbital Hall conductivity ($\sigma_{OHE}$) (9,11,16). Experiments have shown these materials have a large $\xi_{OT}$ (~ 0.78) (20) and a long orbital diffusion length ($\lambda_{OHE}$) (~ 60 nm) (16). However, the options for FM materials capable of efficiently converting orbital current to spin current are limited, particularly for perpendicular magnetic anisotropy (PMA) materials with a strong spin-orbit correlation (15,20,23). Therefore, exploring novel PMA FM materials is crucial for advancing orbitronic devices.

Recent advancements in two-dimensional van der Waals (2D-vdW) materials have led to the experimental confirmation of various 2D-vdW magnetic materials, opening new avenues for innovative room-temperature 2D-vdW spintronic technologies (24-27). Among these advancements, 2D-vdW $Fe_3GaTe_2$ has emerged as a particularly promising FM material, drawing considerable attention for its potential applications in tunnel magnetoresistance (28,29), magnetic skyrmions (30-33), and SOT devices (34-39). This interest is driven by its clean surface, large PMA ($K_u >$ 3.88 $\times 10^5$ J/m$^3$), and high Curie temperature ($T_c > 350$ K) (40). Notably, SOT-driven magnetization switching in $Fe_3GaTe_2$-based heterostructures has been experimentally demonstrated using SHMs with large SOC, such as heavy metals (34-36), topological insulators (37), and topological semimetals (38,39). However, the OT-driven magnetization switching of the $Fe_3GaTe_2$ layer has yet to be observed, and the underlying switching mechanism for 2D-vdW ferromagnets remains elusive.

In this work, we experimentally investigated the OT-driven magnetization switching of the 2D-vdW ferromagnet $Fe_3GaTe_2$ through the Ti OHM and provided a theoretical understanding of the OT-driven magnetization switching mechanism. The $Fe_3GaTe_2$-based heterostructures were fabricated and patterned into Hall bar devices, followed by measuring the current-induced magnetization switching. We found that the Ti OHM enables switching of the $Fe_3GaTe_2$ layer at room temperature in the $Fe_3GaTe_2$/Ti device with a significantly lower switching current density ($J_s \sim 1.6 \times 10^6$ A/cm$^2$). Meanwhile, first-principles calculations were employed to analyze the spin-orbit correlation in $Fe_3GaTe_2$ structures, shedding light on the underlying mechanism of OT-driven magnetization switching of 2D-vdW FM materials. These findings highlight the significance of spin-orbit correlation in the PMA FM layer, providing valuable insights for the design and development of 2D-vdW orbitronic devices.

**Discussion and results**

**Orbital Torque Device Based on 2D-vdW Ferromagnet**

In 2D-vdW FM SOT heterostructures, spin currents can be generated by SHMs [e.g., Pt (34-36), $Bi_{1.1}Sb_{0.9}Te_2S_1$ (37), and $WTe_2$ (39)], and then these spin currents flow into 2D-vdW ferromagnets, exerting a torque that switches their magnetization. The efficiency of the magnetization switching mainly depends on the spin Hall angle ($\theta_{SHE}$) of the SHMs. On the other hand, in 2D-vdW FM OT heterostructures, the switching efficiency depends not only on the orbital Hall angle ($\theta_{OHE}$) of the OHMs but also on the orbital-to-spin conversion coefficient ($\eta_{L\text{-}S}$) of the 2D-vdW ferromagnets. OHMs with high $\sigma_{OHE}$ convert charge current into orbital current via the OHE. The orbital current then flows into the adjacent 2D-vdW FM layer, where it is converted into spin current through strong spin-orbit correlation, generating the torque to switch the magnetization of the 2D-vdW FM layer, as shown in the left panel of Fig. 1a. Therefore, both OHMs with high $\sigma_{OHE}$ and 2D-vdW FM materials with strong spin-orbit correlation are essential for enhancing the $\xi_{OT}$.

Here, we selected Ti as the OHM and the 2D-vdW $Fe_3GaTe_2$ as the PMA FM layer to investigate the charge-to-orbital-to-spin current conversion and the OT-driven magnetization switching. The charge current can efficiently convert into the orbital current through the Ti OHM, which has a larger $\sigma_{OHE}$, calculated to be approximately 4600 ($\hbar/e$)(S/cm). It is worth noting that, due to its weak SOC, Ti exhibits a negligible spin Hall conductivity ($\sigma_{SHE}$) of only about 11 ($\hbar/e$)(S/cm), effectively ruling out the possibility of any significant SOT effect, as shown in Fig. 1b (16). The orbital currents generated in the Ti OHM layer can be efficiently converted into spin currents due to the strong SOC in the $Fe_3GaTe_2$ layer (see the right panel of Fig. 1a).

The $Fe_3GaTe_2$ possesses the hexagonal lattice structure with two adjacent quintuple-layered substructures separated by a vdW gap, where each quintuple layer consists of a $Fe_3Ga$ heterometallic slab sandwiched between two Te layers, as illustrated in Fig. 1a. $Fe_3GaTe_2$ single crystal samples, grown by the self-flux method, exhibit high-quality crystallinity, as demonstrated by prominent (*00L*) Bragg peaks (see Supplementary Fig. 1a) with an estimated Curie temperature of ~ 365 K (see

Supplementary Fig. 1b and Supplementary Note 1 for more details). Figure 1c presents the anomalous Hall resistance ($R_{AHE}$) versus (vs.) out-of-plane external magnetic field ($H_{ext}$) as a function of temperature for the $Fe_3GaTe_2$ nanoflake. The square $R_{xy}$ vs. $H_{ext}$ loops affirm excellent PMA. Meanwhile, the single crystal feature and vdW layered structure were verified by atomic-resolution scanning transmission electron microscopy (STEM) measurements, as shown in Fig. 1d. These results verify that the $Fe_3GaTe_2$ single crystal samples are of high quality, making them suitable for investigating 2D-vdW ferromagnet OT heterostructures.

**Orbital Torque Switching of 2D-vdW Ferromagnet**

To experimentally investigate the OT-driven magnetization switching, we fabricated the 2D-vdW FM $Fe_3GaTe_2$/Ti samples on the Si/SiO$_2$ substrates, which were patterned into Hall bar devices (15 μm × 6 μm) (see the inset of Fig. 2a and Supplementary Fig. 2a) and characterized for current-induced magnetization switching through the OT (see the device fabrication and transport-property measurements in Methods). To assess the PMA of the $Fe_3GaTe_2$ (15.8 nm)/Ti (10.0 nm) Hall bar device, the room temperature $R_{AHE}$ vs. $H_{ext}$ loop was measured with the in-plane $H_{ext}$, as shown in Fig. 2a and Supplementary Fig. 2b. The effective anisotropy fields ($H_K$) were estimated to be 4.02 T ~ 5.33 T as the temperature decreased from 300 K to 225 K, verifying a very strong PMA (36,38,39). Meanwhile, the $R_{AHE}$ vs. $H_{ext}$ loops at even lower temperatures were measured to further characterize the PMA, as plotted in Fig. 2b. The Hall bar device demonstrates 100% remanence with well-defined rectangle $R_{AHE}$ vs. $H_{ext}$ loops. The coercivity ($H_C$) of the Fe$_3$GaTe$_2$ (15.8 nm)/Ti (10.0 nm) Hall bar device increases from 19 mT at 300 K to 540 mT at 10 K, indicating very strong PMA. In addition, as the device was cooled from 300 K to 225 K, the value of $R_{xy}$ increases from 4.7 Ω to 7.7 Ω, as plotted in Fig. 2c, suggesting the excellent magnetic properties of the 2D-vdW Fe$_3$GaTe$_2$ layer.

To demonstrate the OT-driven magnetization switching of the 2D-vdW FM Fe$_3$GaTe$_2$ layer, the $R_{AHE}$ vs. the pulse current ($I_{pulse}$) loops of the Fe$_3$GaTe$_2$ (15.8 nm)/Ti (10.0 nm) Hall bar device were measured with 1 ms write-pulse with a 6s delay followed by read pulses (± 1.0 mA) under in-plane $H_{ext}$ ~ ± 20 mT - ± 150 mT applied along the current direction (see details in Supplementary Note 2). The experimental results are shown in Fig. 2d and Supplementary Figs. 3-5. Although the Curie temperature of the Fe$_3$GaTe$_2$ single-crystal sample is around 365 K (see Supplementary Fig. 1b), it will be slightly lower for the 2D-vdW FM Fe$_3$GaTe$_2$ thin film (36). This will decrease the PMA when the $I_{pulse}$ is applied during the measurement of current-induced magnetization switching. Consequently, $R_{AHE}$ vs. $I_{pulse}$ measurements were carried out at 300 K, 275 K, 250 K, and 225 K to ensure the FM state of Fe$_3$GaTe$_2$ with good PMA while studying the temperature-dependent magnetization switching behavior. Figure 2d presents the $R_{AHE}$ vs. $I_{pulse}$ loops of the Fe$_3$GaTe$_2$ (15.8 nm)/Ti (10.0 nm) Hall bar device measured with temperatures from 300 K to 225 K in the presence of a ± 100 mT in-plane $H_{ext}$. The low resistance and high resistance states were observed at the measured temperatures when switching the $I_{pulse}$ from positive to negative polarities. Reversing the direction of the in-plane $H_{ext}$ also resulted in a reversal of the polarity in the $R_{AHE}$ vs. $I_{pulse}$ loops. Meanwhile, the ratio of magnetization switching exhibited an initial increase followed by a decrease as the in-plane $H_{ext}$ increased, as shown in Supplementary Figs. 3-5. This behavior excludes thermal effects and confirms that the 180° magnetization switching is driven by the OT from the Ti OHM. In addition, the $R_{AHE}$ vs. $I_{pulse}$ loops of the Fe$_3$GaTe$_2$ (15.8 nm)/Ti (10.0 nm) Hall bar device become rectangle (see Fig. 2c) as the measured temperature decreased from 300 K to 225 K, suggesting the complete magnetization switching. Figure 2e shows the switching current density ($J_S$) and switching ratio as functions of temperature for the Fe$_3$GaTe$_2$ (15.8 nm)/Ti (10.0 nm) Hall bar device. It can be observed that the $J_S$ gradually increases from $1.6 \times 10^6$ A/cm$^2$ to $4.8 \times 10^6$ A/cm$^2$ as the temperature decreases, attributed to the enhanced magnetic properties (e.g. PMA) of the Fe$_3$GaTe$_2$ layer as the temperature decreases. Notably, we observed that the switching ratio also increases as the devices are cooled. This enhancement may be attributed to reduced thermal perturbation, which leads to improved interlayer coupling between vdW layers and better alignment of magnetic moments of the Fe$_3$GaTe$_2$.

**Orbital Torque vs. Spin-orbit Torque**

To further investigate the torque efficiency of OT, SOT, and the combined SOT+OT, we designed and prepared two additional samples: the SOT sample Fe$_3$GaTe$_2$ (18.3 nm)/Pt (5.0 nm) and the SOT+OT sample Fe$_3$GaTe$_2$ (16.7 nm)/Pt (2.0 nm)/Ti (10.0 nm). These samples were then patterned into the Hall bar devices (24 μm × 8 μm), all of which exhibit excellent PMA properties, as evidenced by their square $R_{AHE}$ vs. $H_{ext}$ loops. Subsequently, current-induced orbital/spin torque magnetization switching was performed under an in-plane $H_{ext}$ applied along the current direction (see Fig. 3, Supplementary Note 2, and Supplementary Figs. 6-9). Unlike the squared $R_{AHE}$ vs. $H_{ext}$ loops, the $R_{AHE}$ vs. $J_S$ loops do not always exhibit a perfectly squared signal measured at room temperature due to the thermal effect which resulted in magnetic moment disarrangement and PMA decrease. To ensure a fair comparison between different devices, we selected the $R_{AHE}$ vs. $J_S$ loops measured at 275 K under an in-plane $H_{ext}$ = 100 mT, where all devices display near square-shaped $R_{AHE}$ vs. $J_S$ loops. We note that the slight variation in $Fe_3GaTe_2$ thickness across different devices does not significantly affect $J_S$, as confirmed by our test measurements with various thicknesses. Figures 3a-3c present the $R_{AHE}$ vs. $J_S$ loops measured at 275K of $Fe_3GaTe_2$ (15.8 nm)/Ti (10.0 nm), $Fe_3GaTe_2$ (16.7 nm)/Pt (2.0 nm)/Ti (10.0 nm), and $Fe_3GaTe_2$ (18.3 nm)/Pt (5.0 nm) Hall bar devices. Deterministic magnetization switching was observed in all devices, with $J_S$ estimated to be approximately $2.4 \times 10^6$ A/cm$^2$, $5.9 \times 10^6$ A/cm$^2$, and $9.2 \times 10^6$ A/cm$^2$ for the respective devices. Notably, the $J_S$ of the $Fe_3GaTe_2$ (15.8 nm)/Ti (10.0 nm) device is about four times smaller than that of the $Fe_3GaTe_2$ (18.3 nm)/Pt (5.0 nm) device, suggesting that the OT efficiency in the $Fe_3GaTe_2$/Ti heterostructure is higher than the SOT efficiency in the $Fe_3GaTe_2$/Pt heterostructure. For the combined OT and SOT effect, the $J_S$ ($5.9 \times 10^6$ A/cm$^2$) is higher than in the OT-only $Fe_3GaTe_2$/Ti device but lower than in the SOT-only $Fe_3GaTe_2$/Pt device.

Figure 3d illustrates the possible physical mechanism behind the magnetization switching of these devices with different switching efficiency. For the $Fe_3GaTe_2$/Pt device, the charge current directly converts into the spin current in the Pt layer, the switching efficiency mainly depends on the SOT from the Pt SHM ($\sigma_{SHE,Pt}$), as shown in the left panel of Fig. 3d. However, for the $Fe_3GaTe_2$/Ti device, the switching efficiency is primarily influenced by both the OT of the Ti OHM ($\sigma_{OHE,Ti}$) and the spin-orbital correlation strength of the 2D-vdW FM $Fe_3GaTe_2$ layer. The charge current first converts into the orbital current in the Ti layer and then the orbital current is converted into the spin current via the 2D-vdW FM $Fe_3GaTe_2$ layer, as presented in the right panel of Fig. 3. Considering the large OHE of the Ti, the significantly lower $J_S$ of the $Fe_3GaTe_2$/Ti device compared to that of the $Fe_3GaTe_2$/Pt device indicates a rather strong spin-orbit correlation within the 2D-vdW FM $Fe_3GaTe_2$ layer. For the $Fe_3GaTe_2$/Pt/Ti device, the Pt SHM serves as an SOT source, and the Ti OHM acts as an OT source. The spin current originates from two mechanisms: the charge-to-spin current conversion through the Pt SHM and the charge-to-orbital-to-spin current conversation through the Ti OHM and the 2D-vdW FM $Fe_3GaTe_2$ layer. During this process, the Pt SHM converts not only the charge current into the spin current but also some of the orbital current from the Ti OHM into the spin current (41,42). The rest orbital current from the Ti OHM will pass through the Pt layer and flow into the 2D-vdW FM $Fe_3GaTe_2$ layer, and then be converted into spin current. The Pt layer may partially screen the orbital current flowing from the Ti OHM into the 2D-vdW FM $Fe_3GaTe_2$ layer, which diminishes the OT switching efficiency compared to that of the pure OT device, $Fe_3GaTe_2$/Ti. As a result, the $J_S$ (~ $5.9 \times 10^6$ A/cm$^2$) of the $Fe_3GaTe_2$/Pt/Ti device falls between the values observed for the $Fe_3GaTe_2$/Ti and the Fe$_3$GaTe$_2$/Pt devices. We also summarize the $J_S$ as a function of the in-plane $H_{ext}$ for the 2D-vdW FM Fe$_3$GaTe$_2$ devices measured at room temperature switched by the different SHMs and OHMs, as plotted in Fig. 3e. Very interestingly, the $J_S$ of the Fe$_3$GaTe$_2$ devices switched via the light material Ti through OT is comparable to that of the Fe$_3$GaTe$_2$ devices driven by topological quantum materials through SOT [e.g. Bi$_{1.1}$Sb$_{0.9}$Te$_2$S$_1$ (37), TaIrTe$_4$ (38), and WTe$_2$ (39)]. This finding highlights the highly efficient conversion process of charge-to-orbital-to-spin currents in the 2D-vdW FM Fe$_3$GaTe$_2$ layer, with potential implications for other 2D-vdW ferromagnets.

**Spin-orbit Correlation in the vdW ferromagnet**

Based on the experimental results, we demonstrated that the Fe$_3$GaTe$_2$/Ti device exhibits high OT efficiency, as evidenced by its low $J_S$, which arises from the contribution of both the Ti OHM and the 2D-vdW FM Fe$_3$GaTe$_2$ layer. The $\sigma_{OHE}$ of the Ti OHM has been calculated to be ~ 4600 ($\hbar/e$)(S/cm) (see Fig. 1b), making it one of the highest among OHMs. The $\eta_{L-S}$, which represents the strength of spin-orbit correlation $\langle L \cdot S \rangle$ in 2D-vdW ferromagnets, is also expected to be significantly stronger. To verify this unique characteristic of the Fe$_3$GaTe$_2$, we chose the monolayer, bilayer, trilayer, and bulk structures for theoretical calculations, as shown in Fig. 4a. Comprehensive electronic simulations based on density functional theory were then performed to determine their spin-orbit correlation functions (see calculation details in the Methods and Supplementary Note 4). The FM ground state of the Fe$_3$GaTe$_2$ was adopted in calculations for all the structures. For the monolayer Fe$_3$GaTe$_2$, the magnetic moments are mainly located on the Fe-I (~ 2.38 μ$_B$) and Fe-II atoms (~ 1.41

μ$_B$), with small opposite contributions from Ga (~ -0.11 μ$_B$) and Te atoms (~ -0.09 μ$_B$) due to hybridization. Similarly, the bilayer, trilayer, and bulk Fe$_3$GaTe$_2$ exhibit a nearly identical distribution of magnetic moments as the monolayer Fe$_3$GaTe$_2$, due to the relatively weak vdW interaction between the layers. Further analysis of the orbital resolved band structure reveals that the majority of electronic states near the Fermi level (E$_F$) are predominantly contributed by the 3$d$ orbitals of Fe atoms, as illustrated in Fig. 4b for the monolayer Fe$_3$GaTe$_2$.

The spin-orbit correlation $\langle \boldsymbol{L} \cdot \boldsymbol{S} \rangle$ of the FM materials describes the conversion efficiency between the orbital ($\boldsymbol{L}$) and the spin ($\boldsymbol{S}$). Therefore, we calculated the band-resolved spin-orbit correlation function $\langle \boldsymbol{L} \cdot \boldsymbol{S} \rangle_{k,n}$ (9,43) and its integrated value, i.e., spin-orbit correlation coefficient $\eta_{L-S} = \sum_n \int f_{k,n} \langle \boldsymbol{L} \cdot \boldsymbol{S} \rangle_{k,n} dV_{k,n}$ after constructing the effective Hamiltonian using the Wannier90 package (44,45) (see Supplementary Note 4), where $f_{k,n} = f(\varepsilon_{k,n})$ represents the Fermi-Dirac distribution of the nth band. The calculated results of $\langle \boldsymbol{L} \cdot \boldsymbol{S} \rangle_{k,n}$ are summarized in Figs. 4c-4f, in which red and blue colored areas denote strong positive and negative correlations, respectively. The positive value means that orbital angular momentum is converted to spin angular momentum in the same direction, and vice versa (46). It is evident that positive/negative correlation hotspots appear near the E$_F$ (e.g., around the $\boldsymbol{K}$ and $\boldsymbol{K1}$ points) corresponding to strong orbital-to-spin conversion efficiency. Further orbital analysis around these hotspots shows a significant hybridization of the 3$d$ orbitals ($d_{xy}$, $d_{x2-y2}$, $d_{z2}$), which leads to the large spin-orbit correlation $\langle \boldsymbol{L} \cdot \boldsymbol{S} \rangle$. This is because the wave function that consists of $d_{xy}$ and $d_{x2-y2}$ gives a larger matrix element's value of the SOC operator $L \cdot S$ (47). Note that these spin-orbit correlation hotspots occur in only one spin channel, which may further enhance the orbital-to-spin conversion during the OT switching process. Similar correlation hotspots can be observed for bilayer, trilayer, and bulk Fe$_3$GaTe$_2$ structures near the E$_F$ (especially near the **K** and **K1** point), as plotted in Figs. 4d-4f, confirming the robust spin-orbit correlation in Fe$_3$GaTe$_2$.

The spin-orbit correlation coefficient $\eta_{L\text{-}S}$ was calculated to be around 0.375, 0.755, 1.153, and 0.762 for the monolayer, bilayer, trilayer, and bulk Fe$_3$GaTe$_2$ structures, respectively. This variation of $\eta_{L\text{-}S}$ is due to the increased energy bands that contribute the $\langle L \cdot S \rangle_{k,n}$ blow the E$_F$ as the number of layers increases. Therefore, to compare the effective orbital-to-spin conversion efficiency in the monolayer, bilayer, trilayer, and bulk Fe$_3$GaTe$_2$ structures, we defined $\eta_{L-S}^{\text{eff}} = \eta_{L-S}/N$ ( N is the number of layers) to assess the spin-orbit correlation of each layer, which was estimated to be around 0.3752, 0.3773, 0.3842, and 0.3809, respectively, almost independent of the layer thickness. It is important to note that the 2D-vdW PMA ferromagnets are fundamentally different from traditional bulk PMA FM material due to their unique 2D-vdW nature. The switching of each magnetic layer could be much easier because of the robust thickness-independent $\eta_{L-S}^{\text{eff}}$ and the rather weak interlayer coupling. These results not only reveal the different orbital-to-spin conversion mechanisms between the 2D-vdW ferromagnets and traditional ferromagnets but also provide insight into understanding the orbital current transport in the ferromagnets with unique 2D-vdW features.

**Discussion**

To summarize, we investigated the current-induced magnetization switching of the 2D-vdW FM $Fe_3GaTe_2$ heterostructures using Ti, Pt, and Pt/Ti. We found that the OT from the Ti OHM enables efficient magnetization switching of the 2D-vdW FM $Fe_3GaTe_2$ at 275 K with the $J_s \sim 2.4 \times 10^6$ A/cm$^2$, compared to the SOT from the Pt ($J_s \sim 9.2 \times 10^6$ A/cm$^2$) and the combined SOT+OT from the Pt/Ti ($J_s \sim 5.9 \times 10^6$ A/cm$^2$). This indicates a highly efficient charge-to-orbital-to-spin current conversion in the 2D-vdW $Fe_3GaTe_2$/Ti heterostructure. The underlying physical mechanism is that the Ti OHM efficiently converts the charge current into the orbital current due to its high $\sigma_{OHE} \sim 4600$ ($\hbar/e$)(S/cm), while the $Fe_3GaTe_2$ effectively converts the orbital current into the spin current via a significant and layer-independent $\eta_{L-S}$, as confirmed by the spin-orbit correlation calculations.

The efficient 2D-vdW orbitronic devices can be realized through the optimal selection of the OHMs and the 2D-vdW ferromagnet to obtain the efficient conversion of charge-to-orbital-to-spin current. Most importantly, investigating the conversion process and transport properties of orbital current in the 2D-vdW FM $Fe_3GaTe_2$ is crucial for understanding the physical mechanisms behind orbital-to-spin current conversion and magnetization switching in 2D-vdW ferromagnet OT heterostructures. In addition, because of the weak interlayer coupling, 2D-vdW FM materials may lead to much faster magnetization switching of each FM layer. Our experimental and theoretical findings carry important implications for the development of efficient 2D-vdW orbitronic memory and logic devices.

**Methods**

**Crystal growth and property characterizations:** High-quality $Fe_3GaTe_2$ single crystal samples were grown by the self-flux method. Tellurium (Te) powder (99.999%), gallium (Ga) balls (99.9999%), and iron (Fe) powder (99.95%) in a molar ratio of 2:1:2 were sealed in an evacuated quartz tube using a hydrogen-oxygen cutting machine. Subsequently, the quartz tube was heated from room temperature to 1000 °C for 60 minutes in a muffle furnace and maintained at 1000 °C for 1440 minutes, then cooled down from 1000 °C to 780 °C with three-temperature steps. After that, the quartz tube was quenched in the ice-water. The crystalline and magnetic properties of the $Fe_3GaTe_2$ flakes were characterized by powder X-ray diffraction (XRD), scanning transmission electron microscopy (STEM), physical property measurement system (PPMS), and Vibrating Sample Magnetometer (Model 3107, East Changing Technologies, China).

**Device fabrication and transport-property measurements:** 2D-vdW FM $Fe_3GaTe_2$/X (X = Ti, Pt, and Pt/Ti) Hall bar devices were prepared by combining mechanical exfoliation and magnetron sputtering. 2D-vdW FM $Fe_3GaTe_2$ layers were exfoliated on $Si/SiO_2$ substrates in an Argon-filled glove box and then transferred into the chamber of the sputtering system with a base pressure lower than $5.0 \times 10^{-8}$ Torr. After slight surface cleaning, the Ti, Pt, and Pt/Ti layers were deposited on the $Fe_3GaTe_2$ layer capped with a 3.0-nm-thick $SiO_2$ layer. During the deposition process, the Ar working pressure is 3.0 mTorr. Subsequently, the $Fe_3GaTe_2$/Ti, $Fe_3GaTe_2$/Pt/Ti, and Fe$_3$GaTe$_2$/Pt samples were patterned into 4-terminal Hall bar devices through the standard photolithography and an Ar ion milling technique.

The anomalous Hall resistance ($R_{AHE}$) vs. the external magnetic field ($H_{ext}$) loops of the Fe$_3$GaTe$_2$/Ti, Fe$_3$GaTe$_2$/Pt/Ti, and Fe$_3$GaTe$_2$/Pt Hall bar devices were measured using the Electrical Transport Option of the PPMS Dynacool system. Current-induced magnetization switching experiments were performed with a fixed $H_{ext}$ of ±20 mT - ±150 mT along the current direction by interfacing a Keithley 6221 current source and 2182A nanovoltmeter in the Multi-Field Technology Company system.

**Theoretical calculation:** Bulk vdW FM Fe$_3$GaTe$_2$ shares hexagonal lattice structure with space group P6$_3$/mmc (a = b = 3.986 Å, c = 16.229 Å, α = β = 90°, γ = 120°). In each Fe$_3$GaTe$_2$ layer, the Fe$_3$Ga heterometallic slab is sandwiched between two Te layers, as illustrated in Fig. 1a. To eliminate interactions between slabs along the z direction, we adopted 20 Å vacuum layer along the *z*-axis. The adjacent slabs were connected by weak vdW forces with an interlayer spacing of 0.81 nm. We performed first-principles calculations based on the density functional theory (DFT) as implemented in the Vienna *ab initio* simulation package (VASP) (48,49), which is treated by the projector-augmented plane-wave (PAW) method and utilized a plane wave basis set (50). The exchange-correlation potential terms were considered at the level of generalized gradient approximation (GGA) within the scheme of Perdew-Burke-Ernzerhof (PBE) functional (51). For few-layer and bulk structures, the long-range vdW interactions [DFT-D3 method (52)] were incorporated to correct its total energy. The plane-wave cutoff energy is chosen as 400 eV, and we sample the Brillouin zone on 15×15×1 and 15×15×3 regular mesh for the self-consistent calculations of few-layer and bulk $Fe_3GaTe_2$ structures, respectively. The geometric optimizations were performed with a convergence criterion of $10^{-5}$ eV.

**Acknowledgments:** This work was supported by the National Key R&D Program of China (2022YFA1204003), the National Natural Science Foundation of China (Grant Nos. 52271240, U23A20551, 12204037) and the Key project of the Natural Science Foundation of Tianjin (Grant No. 23JCZDJC00400). D.L.Z gratefully acknowledges the research funding provided by the Cangzhou Institute of Tiangong University (Grant No. TGCYY-F-0201) and the Key R&D Program of Cangzhou (222104008). We would like to thank the Analytical & Testing Center of Tiangong University.

**Author contributions:** D.L.Z., H.S.W., J.Y.D., and J.L.C. contributed equally to this work. D.L.Z conceived the work and designed all the samples. Y.J. coordinated and supervised the project. H.S.W., and J.Y.D. prepared the samples and patterned the Hall bar devices with the help by Y.H.Y. and J.L.G.. J.Y.D. and H.S.W. carried out the current-induced orbital/spin-torque magnetization switching experiments with the help P.W., S.H., and Z.Y.J.. J.L.C., and W.J. performed the first-principles calculations. D.D.Y., J.X.Y., and K.Z. prepared the $Fe_3GaTe_2$ single crystals and characterized the crystalline structure and magnetic properties. W.H.W. and Y.L. help analyze the experimental data. D.L.Z. and W.J. wrote the manuscript with inputs from all the authors. All the authors discussed the results and commented on the manuscript.

**Competing interests:** Authors declare no competing interests.

**Data availability:** All data are available in the manuscript or the Supplementary Materials.

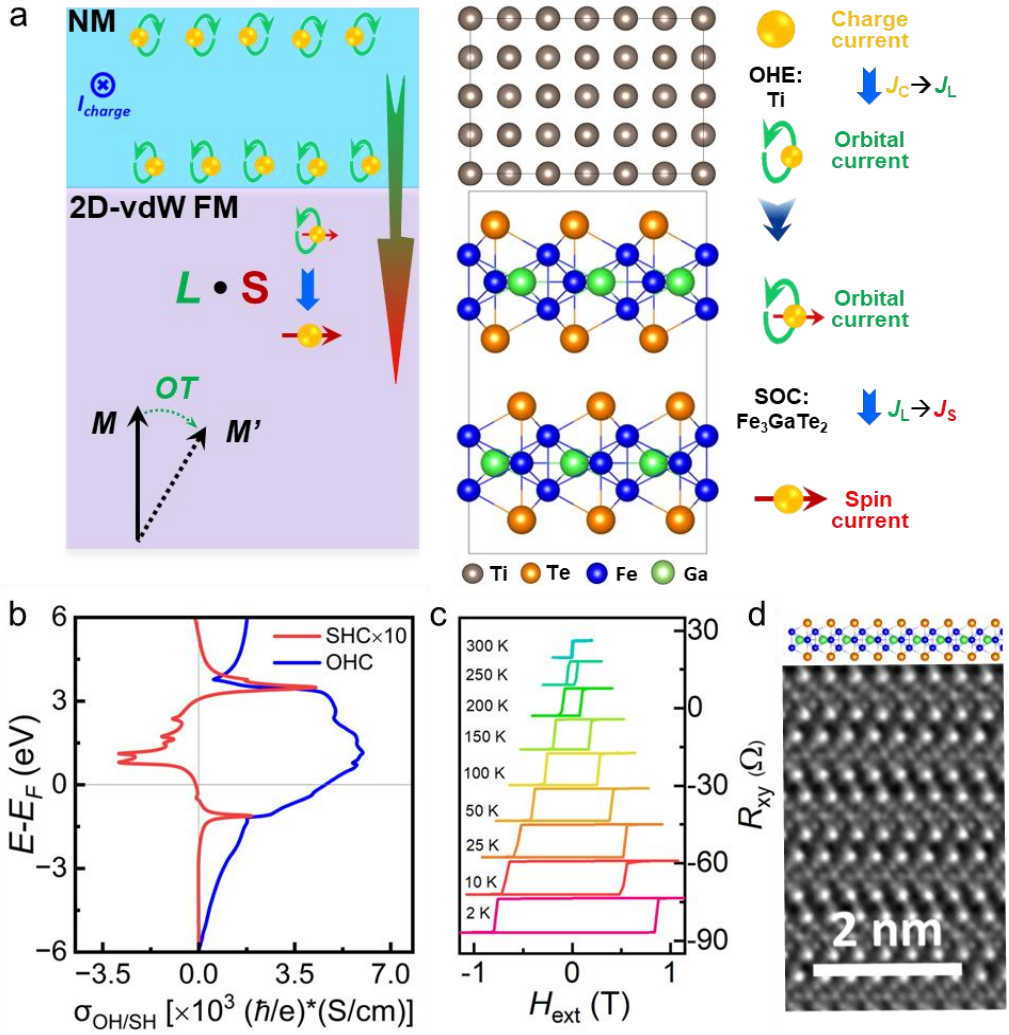

**Fig. 1. Orbital torque in van der Waals ferromagnet. a.** Schematic of the $Fe_3GaTe_2$/Ti orbital torque heterostructure, in which the Ti orbital Hall material converts the charge current ($J_C$) into the orbital current ($J_L$), then the orbital current ($J_L$) flows into the 2D-vdW FM $Fe_3GaTe_2$ layer and is converted into the spin current ($J_S$) due to the spin-orbital coupling of the $Fe_3GaTe_2$ layer. **b.** The calculated orbital Hall conductivity (blue line) and spin Hall conductivity (red line) of the Ti orbital Hall material. **c.** The anomalous Hall resistance ($R_{xy}$) vs. out-of-plane magnetic field $H_{ext}$ as a function of the temperature of the $Fe_3GaTe_2$ nanoflake. **d.** The atomic-resolution HAADF-TEM image of the $Fe_3GaTe_2$ single crystal.

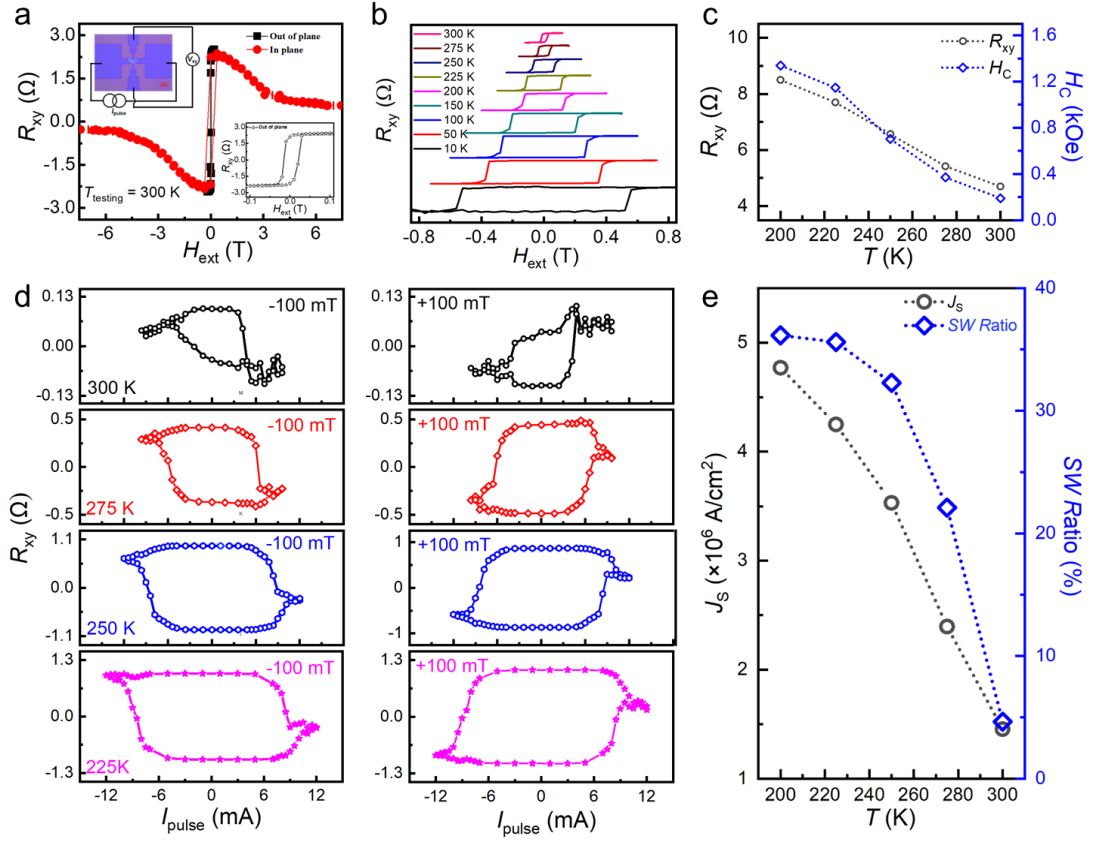

**Fig. 2. Orbital torque switching of 2D vdW ferromagnet. a.** The anomalous Hall resistance ($R_{AHE}$) vs. the external magnetic field ($H_{ext}$) loops measured at room temperature of the 2D-vdW $Fe_3GaTe_2$/Ti Hall bar device, where the $H_{ext}$ is applied along in-plane and out-of-plane directions. The insets show the image of the Hall bar device and the zoom-in out-of-plane $R_{AHE}$ vs. $H_{ext}$ loop. **b.** The $R_{AHE}$ vs. $H_{ext}$ loops as a function of measuring temperatures of the $Fe_3GaTe_2$/Ti device. **c.** The curves of the $R_{AHE}$ vs. $T_{testing}$ and the coercivity ($H_C$) vs. $T_{testing}$ of the $Fe_3GaTe_2$/Ti Hall bar device. **d.** The $R_{AHE}$ vs. the pulse current ($I_{pulse}$) loops of the $Fe_3GaTe_2$/Ti device measured with the in-plane $H_{ext}$ = ± 100 mT along the current direction at temperatures from 300 K to 225 K. **e.** The $J_S$ vs. $T_{testing}$, and the switching ratio vs. $T$ of the $Fe_3GaTe_2$/Ti device.

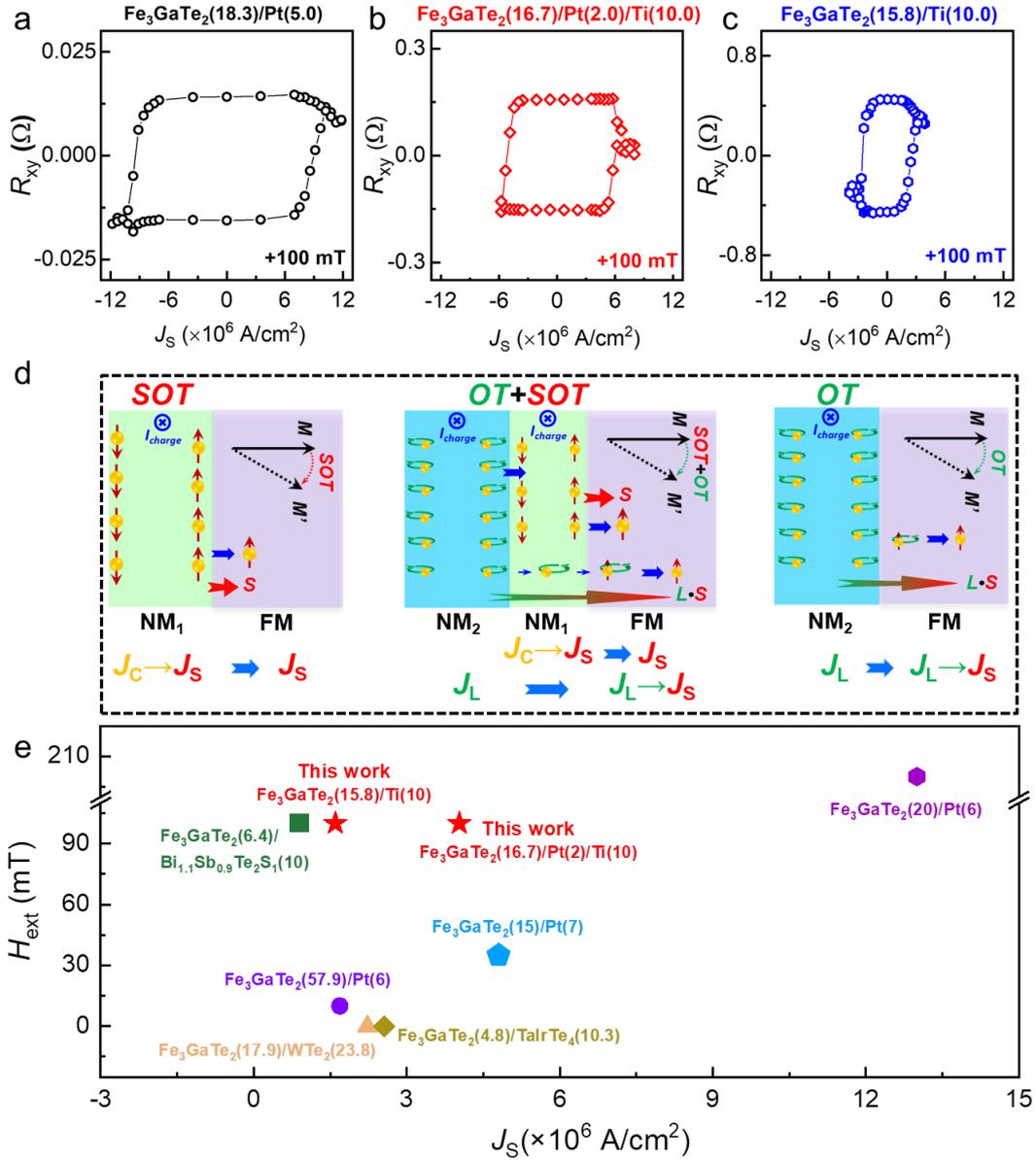

**Fig. 3. Orbital torque vs. Spin-orbit torque. a-c.** The anomalous Hall resistance ($R_{AHE}$) vs. the switching current density ($J_S$) loops measured at 275 K of the 2D-vdW $Fe_3GaTe_2$/Pt, $Fe_3GaTe_2$/Pt/Ti, and $Fe_3GaTe_2$/Ti Hall bar devices, respectively, under the in-plane external magnetic field ($H_{ext}$) = 100 mT along the current direction. **d.** The physical mechanisms of the current conversion for the SOT, SOT+OT, and OT devices. **e.** The summarized $J_S$ values as a function of the in-plane $H_{ext}$ for 2D-vdW $Fe_3GaTe_2$ heterostructures driven through the SOT and OT at room temperature.

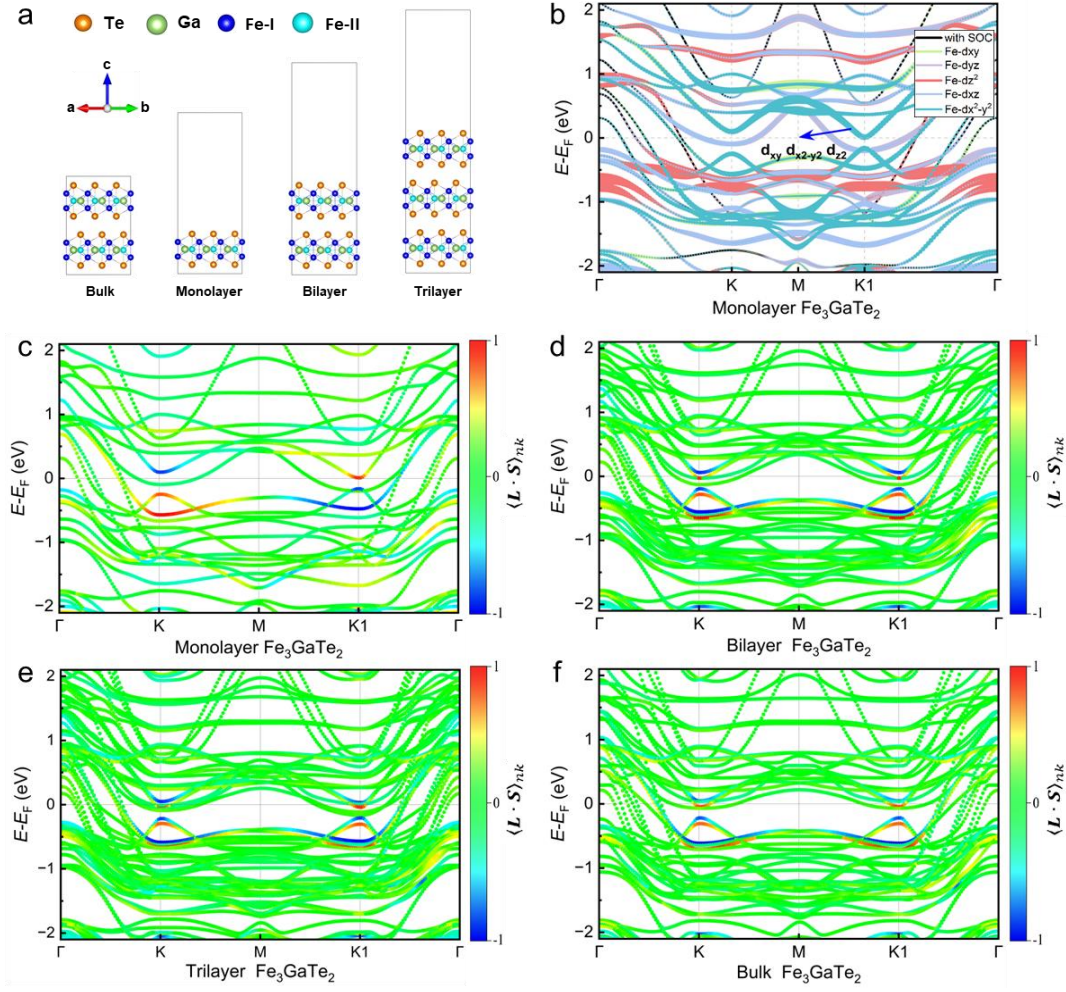

**Fig. 4. Theoretical calculation of spin-orbit correlation function $\langle L \cdot S \rangle_{nk}$. a.** The monolayer, bilayer, trilayer, and bulk $Fe_3GaTe_2$ structures, in which the Fe-I, Fe-II, Ga, and Te atoms are colored blue, light blue, green, and orange, respectively. **b.** The orbital-projected band structure of monolayer $Fe_3GaTe_2$ structure with different $d$ states of Fe is highlighted by different colors. **c-f.** The calculated band-resolved spin-orbit correlation function $\langle L \cdot S \rangle_{nk}$ of the monolayer, bilayer, trilayer, and bulk $Fe_3GaTe_2$ structures, respectively. The color represents the correlation for each eigenstate, in which red and blue denote strong positive and negative correlations, respectively. Note that the Fermi energy is set to zero for reference.